\documentclass[12pt,technote,onecolumn]{IEEEtran}

\usepackage[T1]{fontenc}

\ifCLASSINFOpdf
\else
\fi

\usepackage{amsfonts}
\usepackage{amsthm}
\usepackage{amssymb}
\usepackage{mathtools}
\usepackage{amsmath} 

\theoremstyle{definition}

\usepackage{url} 
\usepackage{blindtext, graphicx}
\usepackage[font=small,skip=0pt]{caption}

\usepackage{float} 
\usepackage{subcaption} 

\usepackage{multirow}
\usepackage{array}

\usepackage[numbers,sort&compress]{natbib}

\usepackage{booktabs}
\usepackage{tabularx}

\usepackage{supertabular}

\usepackage{dblfloatfix}

\usepackage{subcaption} 

\usepackage{enumitem}

\usepackage[linesnumbered,ruled,vlined]{algorithm2e}

\usepackage[table]{xcolor} 

\begin{document}

\title{ADDENDUM \\ Details of the Derivation of the Probability of Coverage for the Relaying Scheme (Section IV in the paper: ``A Poisson Line Process based Framework for Determining the Needed RSU Density and Relaying Hops in Vehicular Networks'')}

\author{Hussein~A.~Ammar, Abdel-karim Ajami
		and~Hassan~Artail\\
		Department of Electrical and Computer Engineering\\
		American University of Beirut, Beirut 1107 2020\\
		e-mail: \{haa141, aa377, hartail\}@aub.edu.lb
\thanks{This document is an addendum for Section IV of the paper~\cite{VehicularRelayingJournal} accepted to be published at the IEEE Transactions on Wireless Communications.}
}


\maketitle
\makeatletter
\def\tagform@#1{\maketag@@@{\normalsize(#1)\@@italiccorr}}
\makeatother


\section{Abstract}
This paper develops a framework to study multi-hop relaying in a vehicular network consisting of vehicles and Road Side Units (RSUs), and the effect of this relaying on the network coverage and the communication delay. We use a stochastic geometry model that consists of a combination of Poisson Line Process (PLP) and 1D Poisson Point Process (PPP) to reliably characterize the vehicular network layout and the locations of the vehicles and the RSUs. Using this model, we analyze the effect of the different network parameters on the coverage provided by the RSUs to the vehicles. Then, we investigate how the uncovered vehicles can receive their intended packets by relaying them through multiple hops that form connected paths to the RSUs. We also analyze the delay introduced to packet delivery due to multi-hop relaying. Namely, we present results that illustrate the coverage gains achieved through multi-hop relaying and the delays induced. Such results could be used by network planners and operators to decide on the different configurations and operational parameters of the vehicular network to suit particular scenarios and objectives.

\section{System Model}
\begin{figure}[H]
	\vspace{-2em}
	\centering
	\includegraphics[width=0.5\textwidth]{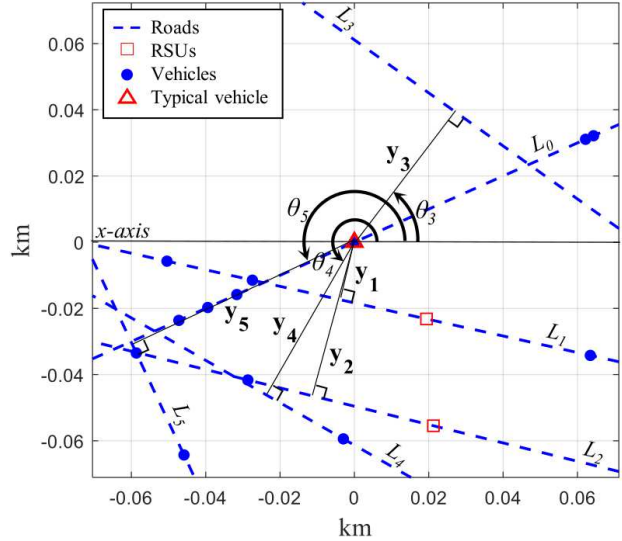}
	\caption{Zoomed view of the vehicular network.}
	\label{fig:NetworkModel1Zoomed}
\end{figure}

We consider a vehicular network that consists of RSUs and vehicles communicating on a typical geographical area containing roads. Since in most areas, the roads' layout is a set of straight and randomly orientated lines, we model these roads' locations as a motion-invariant Poisson Line Process (PLP) $\mathrm{\Phi}_l$~\cite{chiu2013stochastic} with a line density $\rho$. The lines of PLP can be represented by two parameters. The first one is the distance $y_n\in[0,\infty)$ for $n=\{1,2,3,\text{etc}\}$, which represents the perpendicular projection (or simply the distance) from the origin $o$ to  the $n^{th}$ nearest line $L_n$. The second parameter is the angle $\theta_n\in [0, 2\pi)$ between the $x$-axis and $y_n$ measured in a counterclockwise direction (check Figure~\ref{fig:NetworkModel1Zoomed}). When the angles of the lines have a uniform probability measure on $(0, 2\pi]$, the PLP is motion-invariant. We denote the density of the equivalent Poisson Point Processes (PPP) of $\mathrm{\Phi}_l$ on a 2D space $\mathcal{S}=[0,2\pi) \times [0,\infty)$ as $\lambda_l=\frac{\rho}{\pi}$. 
On each road $L_n$, i.e., on each line from the PLP, we use two independent 1D PPP $\mathrm{\Phi}_{{\rm ru}_n}$ and $\mathrm{\Phi}_{{\rm v}_n}$ with respective densities $\lambda_{{\rm ru}}$ and $\lambda_{\rm v}$ to model the RSUs and the vehicles locations, respectively. 
The RSUs are the service providers for the vehicles, i.e., they are the transmitters, and they are connected to each other through backhaul links~\cite{7973038}. Although, we are studying the downlink performance, it is worth mentioning that due to the channel reciprocity and the considered Tx parameters for the RSUs and the vehicles, the uplink performance is the same as the downlink. If we assume that the vehicles are transmitting according to a probability $p_1$, we obtain on each line $n$ a 1D PPP $\mathrm{\Phi}_{{\rm vt}_n}$ with a density $p_1\lambda_{\rm v}$ representing the locations of the vehicles currently transmitting, which follows from the independent thinning on each $\mathrm{\Phi}_{{\rm v}_n}$~\cite{haenggi2012stochastic}. Also, consistent with the literature (e.g.~\cite{6042301}) we assume that the sent signals experience Rayleigh fading with mean $1/\mu$.
\par We study the average network performance at a typical vehicle (observation point) $v_{\rm typ}$ found on the origin $o$, and served by the nearest RSU. According to Slivnyak's Theorem~\cite{haenggi2012stochastic} and the stationarity of the considered model, we can translate the model so that the typical vehicle falls on the origin of the network. The translated model can be treated as the superposition of the PLP with an additional line ($L_0$) passing through the center $\mathrm{\Phi}_{l_0}=\mathrm{\Phi}_l\cup L_0$, and then adding an additional vehicle $v_{\rm typ}$ (the observation point) to the 1D PPP $\mathrm{\Phi}_{{\rm vt}_0}$ falling on this $L_0$~\cite{6260478}\cite{chiu2013stochastic}. We note that by adding $L_0$, we have automatically added $\mathrm{\Phi}_{{\rm ru}_0}$, which is reasonable, else we would be assuming that $v_{\rm typ}$ cannot connect to RSUs on the same road. 
A representation of this model used to create our vehicular network is shown in Figure~\ref{fig:NetworkModel1Zoomed} with a zoomed view of the origin.

\section{Relaying Scheme - (Section IV in manuscript)}
\begin{figure}[H]
	\centering
	\includegraphics[width=0.45\textwidth]{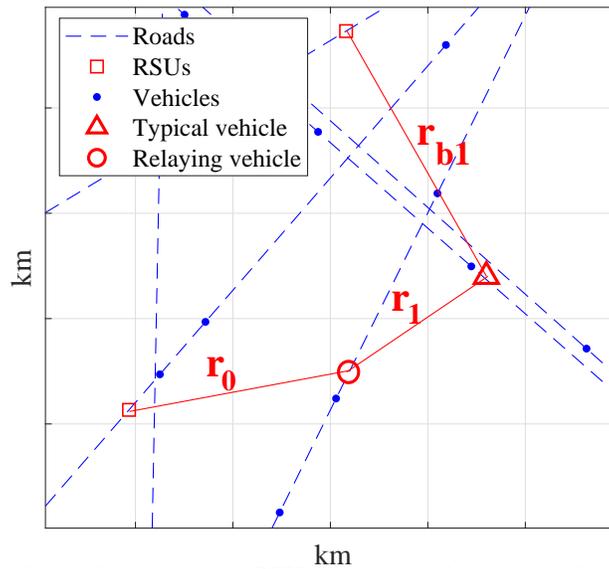}
	\caption{Typical vehicle covered by RSU through a relaying vehicle at distance $r_1$.}
	\label{fig:figureRelaying1}
\end{figure}

We study the scenario when the typical vehicle $v_{\rm typ}$ is not directly covered by its nearest RSU, hence its signal is relayed through another vehicle $v_{{\rm rel}_1}$ found at distance $r_1$, which we denote as Scenario B. In Figure~\ref{fig:figureRelaying1}, we show an example of this scenario.
For this scenario to be successful, $v_{\rm typ}$ must be covered by $v_{{\rm rel}_1}$ given that it is not already covered directly by the serving RSU (its own nearest RSU), and $v_{{\rm rel}_1}$ must be covered by its nearest RSU. The two constraints that render the one-hop relaying scheme useful, are listed below and are shown in Figure~\ref{fig:figureRelaying1}:
\begin{itemize}
	\item $r_{\rm b1}>r_1$: will be accomplished by imposing a lower bound on the variable distance $r_{\rm b1}$.
	\item $r_0<r_{\rm b1}$: will be accomplished by imposing an upper bound on the variable distance $r_0$.
\end{itemize}

\subsection{Summary of the Approach}
We define the Signal to Interference and Noise Ratio (SINR) at the relaying vehicle when it is served by its nearest RSU and the SINR at the typical vehicle when it is served by the relaying vehicle from distance $r_1$. We formulate the probability of coverage for the typical vehicle through one-hop relaying. The formula for this coverage uses the joint Laplace Transform (LT) of interference, which are defined in Lemma 1 in the manuscript and appear naturally in the expression of the coverage because of the dependence of the SINR condition for the typical vehicle before and after relaying, where the typical vehicle will get its signal relayed only when it cannot be directly covered by its nearest RSU. Hence, by using Bayes rule, the joint LT expressions appear.

\subsection{Details of Deriving the Probability of Coverage through One-hop Relaying}
The probability $p_{\rm c}^{{\rm B,rel}_1}$ for a typical vehicle $v_{\rm typ}$ to be covered by an RSU through one-hop relaying is equivalent to the probability that its SINR, received from a relaying vehicle at distance $r_1$, is greater than an SINR threshold $T$; this means that ${\rm SINR}_{\rm B}(r_1) > T$, given that this $v_{\rm typ}$ is not directly covered by the nearest RSU, i.e., ${\rm SINR}_{\rm A}(r_{\rm b1}) < T$. Moreover, the relaying vehicle should be covered by its own nearest RSU, i.e., ${\rm SINR}_{{\rm rel}}(r_0) > T$. Hence we can write the following
\begingroup
\allowdisplaybreaks
\begin{align}
p_{\rm c}^{{\rm B,rel}_1}=\mathbb{P}\left[{\rm SINR}_{\rm B}(r_1)>T|{\rm SINR}_{\rm A}(r_{\rm b1})<T\right]\times\mathbb{P}\left[{\rm SINR}_{{\rm rel}}(r_0)>T\right]
\end{align}
By using Bayes rule for the first term, we express Equation (41) in the manuscript, by writing:
\begin{align}
p_{\rm c}^{{\rm B,rel}_1}&
= \frac{\mathbb{P}\left[{\rm SINR}_{\rm B}(r_1)>T,{\rm SINR}_{\rm A}(r_{\rm b1})<T\right]}{\mathbb{P}\left[{\rm SINR}_{\rm A}(r_{\rm b1})<T\right]} \times\mathbb{P}\left[{\rm SINR}_{{\rm rel}}(r_0)>T\right]
\nonumber \\
&=
\frac{{\xi_1}_{(r_1)}}{{\xi_3}_{(r_1)}} \times {\xi_2}_{(r_1)}
\end{align}
\par Now, we describe how the three elements of Equation (2) are derived. We can therefore think of this section as a detailed version of Appendix C in the manuscript.
\newline
\par Since the small-scale fading experienced by the signal powers is Rayleigh fading, the fading power is exponentially distributed. Hence, we can write the first term ${\xi_1}_{(r_1)}$ as follows \cite[Appendix~A]{6047548}
\begin{flalign}
&{\xi_1}_{(r_1)}=
\resizebox{0.86\columnwidth}{!}
{$\displaystyle
	\mathbb{E}_{r_{\rm b1}}\Bigg[e^{\left(-\frac{\mu T r_1^\eta N}{\nu} - \mu\frac{\kappa Tr_1^\eta}{\nu} I_{{\rm ru}} - \mu Tr_1^\eta I_{{\rm vt}}\right)}\Bigg(1-\bigg(p_{\rm c1}^\text{A}\left( r_{\rm b1},s_7,s_8\right)+\sum_{n=1}^{\infty}\int_{0}^{\infty}p_{\rm c2}^\text{A}\left( r_{\rm b1},s_7,s_8,y_n\right)
	\mathop{}\! \mathrm{d}{y_{n}}\bigg)\Bigg)\Bigg]
	$}
\nonumber\\
&
=
\resizebox{0.86\columnwidth}{!}
{$\displaystyle
	\mathbb{E}_{r_{\rm b1}}\Bigg[e^{\left(-\frac{\mu T r_1^\eta N}{\nu}\right)}L_{I_{{\rm ru}}}\left(s_5\right)L_{I_{{\rm vt}}}\left(s_6\right)\Bigg(1-\bigg(p_{\rm c1}^\text{A}\left( r_{\rm b1},s_7,s_8\right)+\sum_{n=1}^{\infty}\int_{0}^{\infty}p_{\rm c2}^\text{A}\left( r_{\rm b1},s_7,s_8,y_n\right)
	\mathop{}\! \mathrm{d}{y_{n}}\bigg)\Bigg)\Bigg]
	$}
\nonumber&&
\end{flalign}
where $s_5=\frac{\kappa\mu Tr_1^\eta}{\nu}$, $s_6=\mu Tr_1^\eta$, $s_7=\mu Tr_{\rm b1}^\eta$, and $s_8=\frac{\nu\mu Tr_{\rm b1}^\eta}{\kappa}$. Hence
\begin{flalign}
%
%
&{\xi_1}_{(r_1)}
=
\nonumber \\
&
\resizebox{0.97\columnwidth}{!}
{$\displaystyle
	\mathbb{E}_{r_{\rm b1}}\Bigg[e^{\left(-\frac{\mu T r_1^\eta N}{\nu}\right)}L_{I_{{\rm ru}}}\left(s_5\right)L_{I_{{\rm vt}}}\left(s_6\right)-p_{\rm c1}^\text{AB}\left( r_{\rm b1},s_7,s_5,s_8,s_6\right)-\sum_{n=1}^{\infty}\int_{0}^{\infty}p_{\rm c2}^\text{AB}\left( r_{\rm b1},s_7,s_5,s_8,s_6,y_n\right)
	\mathop{}\! \mathrm{d}{y_n}\Bigg]
	$}
\nonumber\\
&=\Bigg[e^{\left(-\frac{\mu T r_1^\eta N}{\nu}\right)}L_{I_{{\rm ru}}}\left(s_5\right)L_{I_{{\rm vt}}}\left(s_6\right)-\int_{r_1}^{\infty}p_{\rm c1}^\text{AB}\left( r_{\rm b1},s_7,s_5,s_8,s_6\right) f_{R}(r_{\rm b1}|\varepsilon_0)
\mathop{}\! \mathrm{d}{r_{\rm b1}}\nonumber\\
&
\quad
-\sum_{n=1}^{\infty}\int_{0}^{\infty}\int_{m(y_n)}^{\infty}p_{\rm c2}^\text{AB}\left( r_{\rm b1},s_7,s_5,s_8,s_6,y_n\right) f_{R}(r_{\rm b1}|\varepsilon_n,y_n)
\mathop{}\! \mathrm{d}{r_{\rm b1}} \mathop{}\! \mathrm{d}{y_n}\Bigg]
\end{flalign}
As can be seen this term involves expectation over the variable $r_{\rm b1}$ with the known Probability Density Functions (PDFs) $f_{R}(r_{\rm b1}|\varepsilon_0)$ and $f_{R}(r_{\rm b1}|\varepsilon_n,y_n)$. Also, we have
\begin{align}\label{eq:pc1AB}
p_{\rm c1}^\text{AB}\left( r_{\rm b1},s_5,s_7,s_6,s_8\right)=&\mathbb{P}\left[\varepsilon_0\right]e^{\left(-\mu TN\left(\frac{r_1^\eta}{\nu}+\frac{r_{\rm b1}^\eta}{\kappa}\right)\right)} L_{\{I_{0},I_{0}\}}\left(\{s_5,s_7\}|\varepsilon_0\right)L_{\{I_{3},I_{3}\}}\left(\{s_5,s_7\}|\varepsilon_0\right)\nonumber\\
&\times L_{\{I_{4},I_{4}\}}\left(\{s_5,s_7\}|\varepsilon_0\right)L_{\{I_{{\rm vt}},I_{{\rm vt}}\}}\left(\{s_6,s_8\}\right)
\end{align}
\vspace{-1.5em}
\begin{align}\label{eq:pc2AB}
&p_{\rm c2}^\text{AB}\left( r_{\rm b1},s_5,s_7,s_6,s_8,y_n\right)=
\resizebox{0.7\columnwidth}{!}
{$\displaystyle
	\mathbb{P}\left[\varepsilon_n|Y_n\right]e^{\left(-\mu TN\left(\frac{r_1^\eta}{\nu}+\frac{r_{\rm b1}^\eta}{\kappa}\right)\right)} L_{\{I_{0},I_{0}\}}\left(\{s_5,s_7\}|\varepsilon_n,y_n\right)L_{\{I_{1},I_{1}\}}\left(\{s_5,s_7\}|\varepsilon_n,y_n\right)
	$}
\nonumber\\
&\ \ \times 
\resizebox{0.95\columnwidth}{!}
{$\displaystyle
	L_{\{I_{2},I_{2}\}}\left(\{s_5,s_7\}|\varepsilon_n,y_n\right)L_{\{I_{3},I_{3}\}}\left(\{s_5,s_7\}|\varepsilon_n,y_n\right) L_{\{I_{4},I_{4}\}}\left(\{s_5,s_7\}|\varepsilon_n,y_n\right)L_{\{I_{{\rm vt}},I_{{\rm vt}}\}}\left(\{s_6,s_8\}\right)f_{Y_n}(y_n)
	$}
\end{align}
The $L_{\{I_{x},I_{y}\}}\left(\{s_5,s_7\}\right)$ are the joint Laplace Transform (LT) terms of $I_x$ and $I_y$ evaluated at $(s_5,s_7)$ and will be discussed later. These expressions appear in the coverage because of the dependence of the SINR condition for the typical vehicle before and after relaying, where the typical vehicle will get its signal relaying only when it cannot be directly covered by its nearest RSU.
\newline
\par The second term ${\xi_2}_{(r_1)}$ can be derived as follows.
\begin{align}\label{xi_2_main}
&{\xi_2}_{(r_1)}=\mathbb{E}_{r_0,r_{\rm b1}}\bigg[p_{\rm c1}^\text{A}\left( r_0,s_3,s_4\right)+\sum_{n=1}^{\infty}\int_{0}^{r_{\rm b1}}p_{\rm c2}^\text{A}\left( r_0,s_3,s_4,y_q\right)
\mathop{}\! \mathrm{d}{y_q}\bigg]
\end{align}
When applying the expectation over the distances $r_0$ and $r_{\rm b1}$, equation \eqref{xi_2_main} can be divided into four cases:
\begin{itemize}
	\item Case 1: Both the relaying and the typical vehicles have their nearest RSU on the same road.
	\item Case 2: The relaying vehicle has its nearest RSU on the same road, and the typical vehicle has its nearest RSU on the $n^\text{th}$ nearest road.
	\item Case 3: The relaying vehicle has its nearest RSU on the $n^\text{th}$ nearest road, and the typical vehicle has its nearest RSU on the same road.
	\item Case 4: The relaying vehicle has its nearest RSU on the $n^\text{th}$ nearest road, and the typical vehicle has its nearest RSU on the $q^\text{th}$ nearest road.
\end{itemize}
We can account for these cases using the following.
\begin{align}
&{\xi_2}_{(r_1)} =\nonumber \\
&
\resizebox{0.97\columnwidth}{!}
{$\displaystyle
	\mathbb{E}_{r_{\rm b1}}\Bigg[\boldsymbol{\Bigg[}\int_{0}^{r_{\rm b1}}p_{\rm c1}^\text{A}\left( r_0,s_3,s_4\right)f_{R}(r_0|\varepsilon_0)
	\mathop{}\! \mathrm{d}{r_0}\boldsymbol{\Bigg]} +\boldsymbol{\Bigg[}\sum_{n=1}^{\infty}\int_{0}^{r_{\rm b1}}\int_{y_q}^{r_{\rm b1}}p_{\rm c2}^\text{A}\left( r_0,s_3,s_4,y_q\right)f_{R}(r_0|\varepsilon_q,y_q)
	\mathop{}\! \mathrm{d}{r_0} \mathop{}\! \mathrm{d}{y_q}\boldsymbol{\Bigg]}\Bigg]
	$}
\nonumber\\
&=\boldsymbol{\Bigg[}
\underbrace{
	\int_{r_1}^{\infty}\bigg(\int_{0}^{r_{\rm b1}}p_{\rm c1}^\text{A}\left( r_0,s_3,s_4\right)f_{R}(r_0|\varepsilon_0)
	\mathop{}\! \mathrm{d}{r_0}\bigg)\mathbb{P}\left[\varepsilon_0\right]f_{R}(r_{\rm b1}|\varepsilon_0)
	\mathop{}\! \mathrm{d}{r_{\rm b1}}
}_{\text{case 1}}
\nonumber\\
&\quad
+
\resizebox{0.945\columnwidth}{!}
{$\displaystyle
	\underbrace{
		\sum_{n=1}^{\infty}\int_{0}^{\infty}\int_{m(y_n)}^{\infty}\bigg(\int_{0}^{r_{\rm b1}}p_{\rm c1}^\text{A}\left( r_0,s_3,s_4\right)f_{R}(r_0|\varepsilon_0)
		\mathop{}\! \mathrm{d}{r_0}\bigg)
		\mathbb{P}\left[\varepsilon_n|Y_n\right]f_{R}(r_{\rm b1}|\varepsilon_n,y_n)f_{Y_n}(y_n)
		\mathop{}\! \mathrm{d}{r_{\rm b1}} \mathop{}\! \mathrm{d}{y_n}
	}_{\text{case 2}}
	\boldsymbol{\Bigg]}
	$}
\nonumber\\
&\quad+
\boldsymbol{\Bigg[}
\underbrace{
	\int_{r_1}^{\infty}\bigg(\sum_{n=1}^{\infty}\int_{0}^{r_{\rm b 1}}\int_{y_q}^{r_{\rm b1}}p_{\rm c2}^\text{A}\left( r_0,s_3,s_4,y_q\right) f_{R}(r_0|\varepsilon_q,y_q)
	\mathop{}\! \mathrm{d}{r_0}
	\mathop{}\! \mathrm{d}{y_q}\bigg)\mathbb{P}\left[\varepsilon_0\right]f_{R}(r_{\rm b1}|\varepsilon_0)
	\mathop{}\! \mathrm{d}{r_{\rm b1}}
}_{\text{case 3}}
\nonumber\\
&\quad
+
\underbrace{
	\sum_{q=1}^{\infty}\int_{0}^{\infty}\int_{m(y_n)}^{\infty}\bigg(\sum_{n=1}^{\infty}\int_{0}^{r_{\rm b 1}}\int_{y_q}^{r_{\rm b1}}  p_{\rm c2}^\text{A}\left( r_0,s_3,s_4,y_q\right)f_{R}(r_0|\varepsilon_q,y_q)
	\mathop{}\! \mathrm{d}{r_0}
	\mathop{}\! \mathrm{d}{y_q}\bigg)\mathbb{P}\left[\varepsilon_n|Y_n\right]
}_{\text{case 4}}
\nonumber\\
&\quad \quad 
\underbrace{
	\times
	f_{R}(r_{\rm b1}|\varepsilon_n,y_n) f_{Y_n}(y_n)
	\mathop{}\! \mathrm{d}{r_{\rm b1}}\mathop{}\! \mathrm{d}{y_n}
}_{\text{case 4 (continued)}}
\boldsymbol{\Bigg]}
\end{align}
where $m(y_n)=\max\{r_1,y_n\}$ is the maximum of $r_1$ and $y_n$. 
The expressions for $p_{\rm c1}^\text{A}$, $p_{\rm c2}^\text{A}$, $\mathbb{P}\left[\varepsilon_0\right]$, $\mathbb{P}\left[\varepsilon_n|Y_n\right]$ $f_{Y_n}(y_n)$ are found in the manuscript. Also, more discussion is found at the end of Appendix C and in Theorem 1.
\newline 
\par As for ${\xi_3}_{(r_1)}$, it is the same probability of outage derived in Scenario A (direct connection to the nearest RSU from the concerned vehicle).
\begin{align}
{\xi_3}_{(r_1)}&=
\resizebox{0.9\columnwidth}{!}
{$\displaystyle
	1-\Bigg[\int_{r_1}^{\infty}p_{\rm c1}^\text{A}\left( r_{\rm b1},s_7,s_8\right)f_{R}(r_{\rm b1}|\varepsilon_0) 
	\mathop{}\! \mathrm{d}{r_{\rm b1}}+\sum_{n=1}^{\infty}\int_{0}^{\infty}\int_{m(y_n)}^{\infty}p_{\rm c2}^\text{A}\left( r_{\rm b1},s_7,s_8,y_n\right)f_{R}(r_{\rm b1}|\varepsilon_n,y_n)
	\mathop{}\! \mathrm{d}{r_{\rm b1}}
	\mathop{}\! \mathrm{d}{y_{n}}\Bigg]
	$}
\end{align}
\endgroup
Now, regarding the joint Laplace transforms (LTs) for the interference found in the equations \eqref{eq:pc1AB} and \eqref{eq:pc2AB} (Lemma 1, equations (32)-(38) in the manuscript) and which depend on the distance $r_{\rm b1}$, there are many works in the literature that show how these LTs can be derived, e.g.,~\cite{6047548} and~\cite{6220221}.

For the case of $L_{\{I_{0},I_{0}\}}\left(\{s_5, s_7\}|\varepsilon_0\right)$, it is derived as follows
\begin{align}
&L_{\{I_{0},I_{0}\}}\left(\{s_5, s_7\}|\varepsilon_0\right)=\mathbb{E}_{g,g',\mathrm{\Phi_{{\rm ru}_0}}}\left[\exp\left(-s_5\sum_{j\in\mathrm{\Phi_{{\rm ru}_0}}}g'_{0,j}R_{0,j}^{-\eta} -s_7\sum_{j\in\mathrm{\Phi_{{\rm ru}_0}}\backslash b(o,r_{\rm b1})}g_{0,j}R_{0,j}^{-\eta} \right)\right]\nonumber\\
&=\mathbb{E}_{g,g',\mathrm{\Phi_{{\rm ru}_0}}}\left[\prod_{j\in\mathrm{\Phi_{{\rm ru}_0}}\backslash b(o,r_{\rm b1})}\exp\left(-s_7g_{0,j}R_{0,j}^{-\eta}\right)\prod_{j\in\mathrm{\Phi_{{\rm ru}_0}}}\exp\left(-s_5g'_{0,j}R_{0,j}^{-\eta}\right)\right]\nonumber\\
&=
\resizebox{0.97\columnwidth}{!}
{$\displaystyle
	\mathbb{E}_{\mathrm{\Phi_{{\rm ru}_0}}}\left[\prod_{j\in\mathrm{\Phi_{ru_0}}\backslash b(o,r_{\rm b1})}\frac{\mu}{\mu+s_7R_j^{-\eta}}\prod_{j\in\mathrm{\Phi_{{\rm ru}_0}}}\frac{\mu}{\mu+s_5R_j^{-\eta}}\right]\stackrel{(a)}{=}\exp\Bigg[-2\lambda_{{\rm ru}}\int_{r_{\rm b1}}^{\infty}\left(1-\frac{\mu}{\mu+s_7x^{-\eta}}\frac{\mu}{\mu+s_5x^{-\eta}}\right)
	\mathop{}\! \mathrm{d}x\Bigg]
	$}
\end{align}
where $r_{\rm b1}>r_1$ and (a) follow from the Probability Generating Functional (PGFL) of the 1D PPP~\cite{haenggi2012stochastic}. The interference from $\mathrm{\Phi_{ru_0}}$ is when the relaying vehicle is transmitting to $v_{\rm typ}$, while the interference from $\mathrm{\Phi_{{\rm ru}_0}}\backslash b(o,r_{\rm b1})$ is when the nearest RSU (w.r.t. to $v_{\rm typ}$) would be transmitting to $v_{\rm typ}$. Additionally, although the term that contains $s_5$ does not contain an exclusion region for the interference (because when the relaying vehicle is sending the data to $v_{\rm typ}$, all RSUs can be an interfering source), we do not add additional terms for the PGFL integral, i.e., $\exp\Bigg[-2\lambda_{{\rm ru}}\bigg(\int_{r_{\rm b1}}^{\infty}\left(1-\frac{\mu}{\mu+s_7x^{-\eta}}\frac{\mu}{\mu+s_5x^{-\eta}}\right)
\mathop{}\! \mathrm{d}x+B\bigg)\Bigg]$, because by definition we already know that the interfering RSUs 
are further than distance $r_{\rm b1}$, and they are already included in the integral.

For the other LTs we follow the same derivations done in~\cite[Lemma 12, 13, 14 and 15]{8340239}, which use the main properties of the 1D PPP and the PLP~\cite{haenggi2012stochastic} and include conditioning on the serving distance $r_{\rm b1}$, the distance $y_n$ from the typical vehicle to the roads, and the event $\epsilon_n$. These derivations are however done with replacing the terms $\frac{s_1}{\mu\left(x^2+y^2\right)^\frac{\eta}{2}+s_1}$ in the LTs from Scenario A in the manuscript by
\begin{align}
\zeta_2\left(x,y_n,s_5,s_7\right)=1-\frac{\mu}{\mu+s_5\left(x^2+{y_n}^2\right)^\frac{-\eta}{2}} \left( \frac{\mu}{\mu+s_7\left(x^2+{y_n}^2\right)^\frac{-\eta}{2}} \right)
\end{align}
The readers can refer to~\cite{8340239} for the detailed derivations mentioned above.

\subsection{Final Notes}
We do not impose on the nearest RSU to be the same for the typical vehicle $v_{\rm typ}$ and for the relaying vehicle $v_{{\rm rel}}$, thus making $r_0$ and $r_{\rm b1}$ independent. Note that by using the properties of the homogeneous Poisson process, the numbers of points $N$ in two disjoint regions $R_1$ and $R_2$ are independent~\cite{haenggi2012stochastic}, and we can directly relate the number of points found in a region to the serving distance through the relation
\begin{align}
\mathbb{P}(\text{dist}(u,\Phi) \le r) = \mathbb{P}(N(b(u,r))>0)
\end{align}
where $\text{dist}(u,\Phi) \triangleq \min\{x \in \Phi: \|x - u\|\}$, with $u \in \mathbb{R}^2$, is the distance from point $u$ to the nearest point in the Poisson process $\Phi$, and $N(b(u,r))$ is the number of points in the circular region of center $u$ and radius $r$.
\par If the nearest RSU to $v_{\rm typ}$ cannot provide the coverage condition, then $v_{\rm typ}$ may get its signal relayed via another vehicle $v_{{\rm rel}}$ that is covered by its corresponding nearest RSU. Hence, it follows that the conditions for the distances $r_0$ and $r_{\rm b1}$ may be from different RSUs. As a result, we can apply the expectations over these distances independently, again, because we do not require that the nearest RSU to the two vehicles to be the same. Our postulation makes the system more flexible and the analysis more tractable as compared to a condition that states that the serving and nearest-to-relay RSUs should be the same. By more flexible we mean that with our design choice, we enable another RSU to send the typical vehicle the intended message. This is a valid argument since the RSUs are considered to be connected, and moreover, via high capacity backhaul links, as stated in several works, like~\cite{7973038}. Hence if the serving RSU cannot deliver the message to $v_{\rm typ}$, it can forward it via the backhaul link to the nearest RSU to $v_{\rm rel}$ that can relay the message to the $v_{\rm typ}$. Not to mention the caching capabilities at the RSUs for the vehicles data, and the fact that data generated by some common services, like map-updates, is likely to be available on all RSUs.

\ifCLASSOPTIONcaptionsoff
  \newpage
\fi

\footnotesize
\bibliography{Model_References}
\bibliographystyle{unsrt}


\end{document}